RESEARCH                                                                                    Open Access

# Towards a HPC-oriented parallel implementation of a learning algorithm for bioinformatics applications

Gianni D'Angelo[1,2], Salvatore Rampone[1,2*]

## Abstract

**Background:** The huge quantity of data produced in Biomedical research needs sophisticated algorithmic methodologies for its storage, analysis, and processing. High Performance Computing (HPC) appears as a magic bullet in this challenge. However, several hard to solve parallelization and load balancing problems arise in this context. Here we discuss the HPC-oriented implementation of a general purpose learning algorithm, originally conceived for DNA analysis and recently extended to treat uncertainty on data (U-BRAIN). The U-BRAIN algorithm is a learning algorithm that finds a Boolean formula in disjunctive normal form (DNF), of approximately minimum complexity, that is consistent with a set of data (instances) which may have missing bits. The conjunctive terms of the formula are computed in an iterative way by identifying, from the given data, a family of sets of conditions that must be satisfied by all the positive instances and violated by all the negative ones; such conditions allow the computation of a set of coefficients (relevances) for each attribute (literal), that form a probability distribution, allowing the selection of the term literals. The great versatility that characterizes it, makes U-BRAIN applicable in many of the fields in which there are data to be analyzed. However the memory and the execution time required by the running are of $O(n^3)$ and of $O(n^5)$ order, respectively, and so, the algorithm is unaffordable for huge data sets.

**Results:** We find mathematical and programming solutions able to lead us towards the implementation of the algorithm U-BRAIN on parallel computers. First we give a Dynamic Programming model of the U-BRAIN algorithm, then we minimize the representation of the relevances. When the data are of great size we are forced to use the mass memory, and depending on where the data are actually stored, the access times can be quite different. According to the evaluation of algorithmic efficiency based on the *Disk Model*, in order to reduce the costs of the communications between different memories (RAM, Cache, Mass, Virtual) and to achieve efficient I/O performance, we design a mass storage structure able to access its data with a high degree of *temporal* and *spatial locality*. Then we develop a parallel implementation of the algorithm. We model it as a SPMD system together to a Message-Passing Programming Paradigm. Here, we adopt the high-level message-passing systems MPI (Message Passing Interface) in the version for the Java programming language, MPJ. The parallel processing is organized into four stages: partitioning, communication, agglomeration and mapping. The decomposition of the U-BRAIN algorithm determines the necessity of a communication protocol design among the processors involved. Efficient synchronization design is also discussed.

**Conclusions:** In the context of a collaboration between public and private institutions, the parallel model of U-BRAIN has been implemented and tested on the INTEL XEON E7xxx and E5xxx family of the CRESCO structure of Italian National Agency for New Technologies, Energy and Sustainable Economic Development (ENEA), developed in the framework of the European Grid Infrastructure (EGI), a series of efforts to provide access to high-throughput computing resources across Europe using grid computing techniques. The implementation is able to minimize

* Correspondence: rampone@unisannio.it
[1]Department of Science and Technology (DST), University of Sannio, Benevento, Italy
Full list of author information is available at the end of the article both the memory space and the execution time. The test data used in this study are IPDATA (Irvine Primate splice-junction DATA set), a subset of HS3D (Homo Sapiens Splice Sites Dataset) and a subset of COSMIC (the Catalogue of Somatic Mutations in Cancer). The execution time and the speed-up on IPDATA reach the best values within about 90 processors. Then the parallelization advantage is balanced by the greater cost of non-local communications between the processors. A similar behaviour is evident on HS3D, but at a greater number of processors, so evidencing the direct relationship between data size and parallelization gain. This behaviour is confirmed on COSMIC. Overall, the results obtained show that the parallel version is up to 30 times faster than the serial one.

## Background

The huge quantity of data produced in Biomedical research needs sophisticated algorithmic methodologies for its storage, analysis, and processing [1]. Examples of the huge databases available throughout the world are given in the annual Database Issue of Nucleic Acids Research and in the on line Molecular Biology Database Collection [2-4].

Furthermore in many applications one must deal with data that have been collected incompletely [5,6]. For example in medical studies, measurements on some subjects may be partially lost at certain stages of the treatment [7]; in DNA analysis, gene- expression microarrays may be incomplete due to insufficient resolution, image corruption, or simply dust or scratches on the slide [8]; in sensing applications, a subset of sensors may be absent or fail to operate at certain regions [9].

Incomplete data problems are often solved by filling the missing data with specific values (imputation method). Common algorithms that have been used to complete missing data include: semidefinite programming [10], the EM algorithm [11,12], Naïve Bayes classifiers [13], C4.5 [14], Gibbs sampling [15], gradient descent [16]. Since these methods rely on the assumption that data are Missing at Random (MAR) [17] or they treat the missing data as fixed known data [18], they suffer of dramatic decrease in accuracy. A full discussion can be found in [17-21].

Along with the growth of the data and the need for solutions in the problem of missing data, there is a great necessity of computationally efficient and scalable algorithms able to extract useful information from data sets of very large size [22-28]. This is one of the main challenges in computational biology, since the tools and the methods capable of transforming the heterogeneous available data into biological knowledge [29] must be implemented efficiently and effectively on the available computer systems.

Recently, in order to deal with incomplete training data, a machine learning algorithm, BRAIN (Batch Relevance-based Artificial INtelligence) [30], for binary classification rules has been generalized (U-BRAIN) [31]. This algorithm was originally conceived for recognizing splice junctions in human DNA (see also [32,33]). Splice junctions are points on a DNA sequence at which "superfluous" DNA is removed during the process of protein synthesis in higher organisms [34]. The general method used in the algorithm is related to the STAR technique of Michalski [35], to the candidate-elimination method introduced by Mitchell [36], and to the work of Haussler [37]. The BRAIN algorithm was then extended by using fuzzy sets [38], in order to infer a DNF formula that is consistent with a given set of data which may have missing bits. The new algorithm (U-BRAIN) has low error rates and keeps the polynomial computational complexity of the original BRAIN algorithm.

Unfortunately the algorithm computational complexity, while polynomial, is unaffordable for large scale data. In fact, the algorithm is based on time-consuming nested cycles that need a lot of memory space to store partial results.

### Overview of the U-BRAIN algorithm

The U-BRAIN algorithm [31] is a learning algorithm that finds a Boolean formula ($f$) in disjunctive normal form (*DNF*) [39], of approximately minimum complexity, that is consistent with a set of data (instances). The conjunctive terms of the formula are computed in an iterative way by identifying, from the given data, a family of sets of conditions that must be satisfied by all the positive instances and violated by all the negative ones; such conditions allow the computation of a set of coefficients (relevances) for each attribute (literal), that form a probability distribution, allowing the selection of the term literals.

The given instances are vectors of $n$ variables. The instances for which $f$ gives the value 1

$$u_1, u_2, ..., u_p \qquad (1)$$

where for each $u_i$

$$u_{ik} \in \{0, 1, 1/2\}, k = 1, ..., n \qquad (2)$$

(1/2 means an uncertain value)
are called positive, while those for which $f$ gives 0

$$v_1, v_2, ..., v_q \qquad (3)$$

where for each $v_j$

$$v_{jk} \in \{0, 1, 1/2\}, k = 1, ..., n \qquad (4)$$

are called negative [31].
We denote

n the number of variables.
2n the number of literals (n in true and n in negated form)
p the number of positive instances.
q the number of negative instances.
i the index of positive instances, ranging from 1 to p.
j the index of negative instances, ranging from 1 to q.

In order to build a formula consistent with the given data, U-BRAIN compares each given positive instance with each negative one and builds a family of sets $S_{ij}$ of literals, each representing a condition:

$$S_{ij} = \left\{x_k | (u_{ik} > v_{jk}) or \left(u_{ik} = v_{jk} = \frac{1}{2}\right)\right\} \cup \left\{\bar{x}_k | (u_{ik} < v_{jk}) or \left(u_{ik} = v_{jk} = \frac{1}{2}\right)\right\} \quad (5)$$

The k-th literal is present in the $S_{ij}$ set if the elements in the position k, belonging to the i-th positive instance ($u_{ik}$) and to the j-th negative instance ($v_{jk}$), are different or both equal to 1/2.

Depending on the type of pair ($u_{ik}$, $v_{jk}$) the literal is taken in true ($x_k$) or negated form ($\bar{x}_k$). In the following a generic literal will be signed $l_k$ for $x_k$ and $l_{n+k}$ for $\bar{x}_k$.

For each literal belonging to each $S_{ij}$ set a relevance $R_{ij}$ is computed as follows:

$$R_{ij}(l_k) = \frac{\chi_{ij}(l_k)}{\#S_{ij}}; \qquad \#S_{ij} = \sum_{m=1}^{2n} \chi_{ij}(l_m) \quad (6)$$

Where $\chi$ is a membership function, defined as:

$$\chi_{ij}(x_k) = \begin{cases} 1 & \text{if } u_{ik} = 1 \text{ and } v_{jk} = 0 \\ \left(\frac{1}{2}\right)^{(p+q)} & \text{if } u_{ik} > v_{jk} \text{ and } (u_{ik} = \frac{1}{2} \text{ or } v_{jk} = \frac{1}{2}) \\ \left(\frac{1}{2}\right)^{(p+q+1)} & \text{if } u_{ik} = \frac{1}{2} \text{ and } v_{jk} = \frac{1}{2} \\ 0 & \text{otherwise} \end{cases} \quad (7)$$

for literals in true form, and

$$\chi_{ij}(\bar{x}_k) = \begin{cases} 1 & \text{if } u_{ik} = 0 \text{ and } v_{jk} = 1 \\ \left(\frac{1}{2}\right)^{(p+q)} & \text{if } u_{ik} < v_{jk} \text{ and } (u_{ik} = \frac{1}{2} \text{ or } v_{jk} = \frac{1}{2}) \\ \left(\frac{1}{2}\right)^{(p+q+1)} & \text{if } u_{ik} = \frac{1}{2} \text{ and } v_{jk} = \frac{1}{2} \\ 0 & \text{otherwise} \end{cases} \quad (8)$$

for literals in negated form.
Then, for each fixed i-th positive instance the $R_i$ relevance is calculated:

$$R_i(l_k) = \frac{1}{q} \sum_{j=1}^{q} R_{ij}(l_k) \quad (9)$$

Finally, the overall $R$ relevance for each literal ranging from 1 to 2n is computed as it follows:

$$R(l_k) = \frac{1}{p} \sum_{i=1}^{p} R_i(l_k) \quad (10)$$

$R(l_k)$ is a 2n dimensional vector in which each element represents a probability value:

$$\sum_{k=1}^{2n} R(1_k) = 1; \quad \text{with } R(l_k) \geq 0 \;\forall\; k \quad (11)$$

The literal $l_k$ having maximum relevance value is chosen as the next literal of the term of the function f.

After the literal choice, the sets (5) are updated: all the $S_{ij}$'s including $l_k$ (satisfied condition) are erased as the $S_{ij}$'s belonging to

$$\{S_{ij} | l_k \notin S_{ij} \text{ for } j = 1 \ldots q\} \quad (12)$$

The cycle is then repeated and the term is completed when there are no more elements in the $S_{ij}$ sets or there are no more $S_{ij}$ sets. Finally the term is added to the function *f*. Then the process starts again after erasing from the given data (1) the positive instances satisfying the term found, and updating the uncertain values and the instances. This last step is very important, since each time a term is produced, the implicit choices over the uncertain components of the negative instances, if any, must be explicated to avoid contradiction with the terms to be generated from now on. Moreover, it is possible that there are some instances that are repeated one or more times, either since the beginning or as a result of the reduction step. The results of this updating phase are checked by a consistency test.

The algorithm ends when there are no more data to treat.

The following is the U-BRAIN algorithm schema.

1. Initialize f = Ø
2. While(∃ positive instances)
   2.1. Uncertainty Reduction
   2.2. Repetition Deletion
   2.3. Initialize term = Ø
   2.4. Build $S_{ij}$ sets
   2.5. While(∃ elements in $S_{ij}$)
      2.5.1. Compute the $R_{ij}$ relevances
      2.5.2. Compute the $R_i$ relevances
      2.5.3. Compute the R relevances
      2.5.4. Choose Literal
      2.5.5. Update term
      2.5.6. Update $S_{ij}$ sets
   2.6. Add term to f
   2.7. Update positive instances
   2.8. Update negative instances
   2.9. Check consistency

### Algorithm complexity
The algorithm complexity refers to both the amount of memory it requires to run to completion (space complexity)

and the amount of time it needs to run to completion (time complexity) [40]. According to the Landau's symbol [41], in the following a big O notation will be used to describe the upper bound complexity.

In order to build a family of sets $S_{ij}$ and to calculate the $R_{ij}(l_k)$ elements, U-BRAIN compares each given positive instance with each negative one.

The cardinality of each $S_{ij}$ is at most 2n, since there are n literals in true and n in negated form. So $R_{ij}(l_k)$ is also valued on 2n literals. This means that the dimensions of $\{S_{ij}\}$ and $\{R_{ij}(l_k)\}$ are as it follows:

$$\#(\{S_{ij}\}) = p \times q \tag{13}$$

$$\#(\{R_{ij}(l_k)\}) = p \times q \times 2n \tag{14}$$

So the space complexity is in the order of $O(pqn) \approx O(n^3)$ for large *n*.

Since each element of a $R_{ij}$ vector is an element of a probability distribution, it is represented by a floating-point number, which, depending on the coding, occupies several Bytes in a computer's internal memory. Thus, storing of $R_{ij}$ for large scale data in a computer memory is space consuming.

*Example 1*: Using the Java language and a data set having

p = 2000, q = 3000, n = 560

the $R_{ij}$ and $S_{ij}$ dimensions are about 430 GByte and 200 GByte respectively for the first iteration.

From the time point of view, since the *external cycle* (2.) is iterated at most p times, the *internal cycle* (2.5.) is iterated at most n times (the maximum length of a term), the inner relevance computation (2.5.1.-2.5.3.) and the $S_{ij}$ update (2.5.6.) are both of $O(pqn)$, and all the other operations are minorities of these, the overall algorithm time complexity is $O(p^2qn^2) \approx O(n^5)$ for large *n*.

## Methods
We find mathematical and programming solutions able to effectively implement the algorithm U-BRAIN on parallel computers. First we give a Dynamic Programming model [42] of the U-BRAIN algorithm; then we minimize the representation of the relevances; finally, in order to reduce the communication costs between different memories and, then, to achieve efficient I/O performance, a mass storage structure is designed to access its data with a high degree of *temporal* and *spatial locality* [43]. Then a parallel implementation of the algorithm is developed by a Single Program Multiple Data (SPMD) technique together to a Message-Passing Programming paradigm.

### Dynamic programming model of the U-BRAIN algorithm
In the U-BRAIN algorithm the $S_{ij}$ sets are built in the external cycle (2.4.), and then, for each resulting $S_{ij}$ set, the $R_{ij}$ relevance vectors are calculated (2.5.1.-2.5.3.). Thus, two memory areas are required, one for the $S_{ij}$ sets and one for the relevance vectors. Moreover, for each choice of a literal, a $S_{ij}$ updating step is done (2.5.6.), consisting in a reduction (erasing) of some $S_{ij}$ sets. Starting from here, the inner cycle (2.5.) is repeated and the new $R_{ij}$ relevance vectors, based on the new $S_{ij}$ sets, are calculated, again. This last step is repeated until there are no more $S_{ij}$ sets corresponding to the production of the term.

It is worth to note that for each inner iteration, the $S_{ij}$ sets are not modified but only erased, and so the recalculation of the $R_{ij}$ relevances on the survivor sets it is not necessary because the new $R_{ij}$ relevances are equal to the ones calculated on the first cycle.

Moreover, it is easy to see that it is possible to calculate the $R_{ij}$ relevance vectors directly from the given data (1) and (2) without using the $S_{ij}$ sets. Then, the $S_{ij}$ sets are unnecessary. Therefore, the $R_{ij}$ relevance vectors can be calculated only once (before and out of the inner cycle) reused and, in case, erased at each inner iteration. So, we modify the U-BRAIN algorithm as follows:

1. Initialize f = Ø
2. While(∃ positive instances)
   2.1. Uncertainty Reduction
   2.2. Repetition Deletion
   2.3. Initialize term = Ø
   **2.4. Compute the Rij relevances**
   2.5. While(∃ elements in **Rij**)
      2.5.1. Compute the *Ri* relevances
      2.5.2. Compute the *R* relevances
      2.5.3. Choose Literal
      2.5.4. Update term
      **2.5.5. Update Rij**
   2.6. Add term to f
   2.7. Update positive instances
   2.8. Update negative instances
   2.9. Check consistency

These changes involve a dramatic reduction in both space of memory, by avoiding the $S_{ij}$ computation, and execution time, by avoiding the $R_{ij}$ computation in the inner cycle.

### Minimizing the Rij representation
Since the $R_{ij}$ relevances are valued for each of the 2n literals as floating-point numbers, a lot of memory space is required to represent them during the algorithm inner cycle execution. Then, a reduced representation form for the $R_{ij}$'s is desirable.

Now each $R_{ij}$ is given by (6), (7) and (8). Aim of the $(\frac{1}{2})^{(p+q+1)}$ values in (7) and (8) is to represent the very low probability of the uncertain literals. This can be also obtained through the following replacement:

$$\chi_{ij}(l_k) = \begin{cases} 1 \\ \left(\frac{1}{2}\right)^{(\alpha)} \\ \left(\frac{1}{4}\right)^{(\alpha)} \\ 0 \end{cases} \text{with } \alpha = p + q. \quad (15)$$

Now, let:

$$\beta_{ij}(l_k) = \begin{cases} 0 \\ 1 \\ \frac{1}{2} \\ \frac{1}{4} \end{cases} \quad (16)$$

Then:

$$\chi_{ij}(l_k) = \beta_{ij}^{\alpha}(l_k) \quad (17)$$

and $R_{ij}$ becomes:

$$R_{ij}(l_k) = \frac{\beta_{ij}^{\alpha}(l_k)}{\sum_{m=1}^{2n} \beta_{ij}^{\alpha}(l_m)} \quad (18)$$

In this way it is possible to view the $R_{ij}$ relevances as function of the four valued coefficients $\beta_{ij}$, each one representable by 2 bits only. This implies a significant reduction of the required memory space.

*Example 2*: Compared to the previous *Example 1*, for

$$p = 2000, q = 3000, n = 560$$

the $R_{ij}$ dimension is decreased from about 430 GByte to 1,7 GByte.

### $\beta_{ij}$ data structure and storage system

Since the $R_{ij}$ computation relies on $\beta_{ij}$'s, we define a data structure to hold them. The $\beta_{ij}$'s related to the i-th positive instance form the i-th $\beta_i$ set. We represent this set as an array (*inner array*). Each element of the inner array contains a $\beta_{ij}$ vectors of 2n dimension whose elements are the 2 bits representation of a $\beta_{ij}$ value. Then we define an array containing each $\beta_i$, named *outer array*.

A schematic representation of the *βij* Data Structure is shown in Figure 1.

The storage of a such data structure in the computer's internal memory, for large data, is unaffordable. Moreover, if this were possible, the huge dimensions may cause delays in I/O operations. In fact, as largely treated in [43], the access times to the computer's internal memory is usually considered to be constant and independent from the address of the memory cell involved and, then, independent from the involved data size. Unfortunately, this is not true for data larger than the internal memory space, forcing the virtual memory use. Depending on where the data are actually stored, the access times can be quite different. Thus, for massive amounts of data, the communication between levels of memory is often a bottleneck. Here, according to the evaluation of algorithmic efficiency based on the *Disk Model*, performed by Vitter in [43], in order to reduce the costs of the communications between different memories and, then, to achieve efficient I/O performance, a mass storage structure, representing the data structure proposed in Figure 1, has been designed to access its data with a high degree of *temporal* and *spatial locality*, as shown in Figure 2.

The elements of the mass storage structure are in sequence. Each element contains three different typology of data: Data Size representing the dimension in bytes of the data to be stored, a Delete Flag that indicates whether the item has been deleted and, finally, the Data section. Each Data section contains a $\beta i$ set, according with the $\beta_{ij}$ data structure reported in Figure 1.

As shown in the Figure 2, the data are stored in a manner that the $\beta_i$ vectors are close together as much as possible (spatial locality). The mass storage structure has been built-out through a random access file which exchange the data section with the computer's internal data structure, typically a vector, through a stream of bytes, as shown in Figure 3. Furthermore, according to the temporal locality, the $\beta i$ data that are referred repeatedly in the same time interval are in a unique block and sequentially ordered. In this way, it is possible to use the data several times and load them in the internal memory only once. Therefore, the access time to the storage device, usually high, becomes negligible compared to the transfer time between internal memory and storage device.

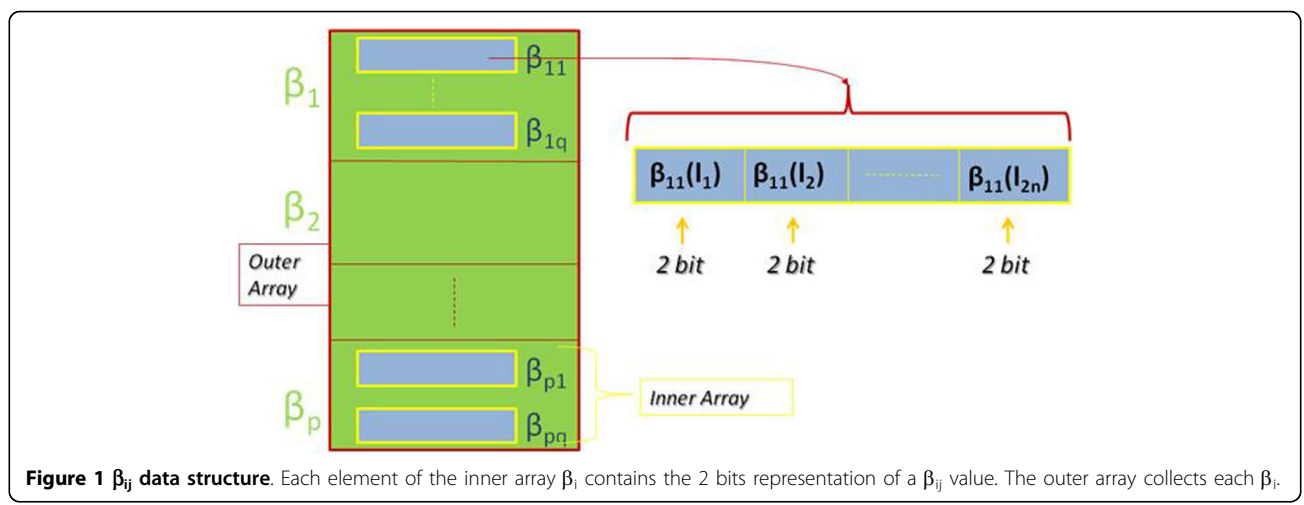

**Figure 1 $\beta_{ij}$ data structure**. Each element of the inner array $\beta_i$ contains the 2 bits representation of a $\beta_{ij}$ value. The outer array collects each $\beta_i$.

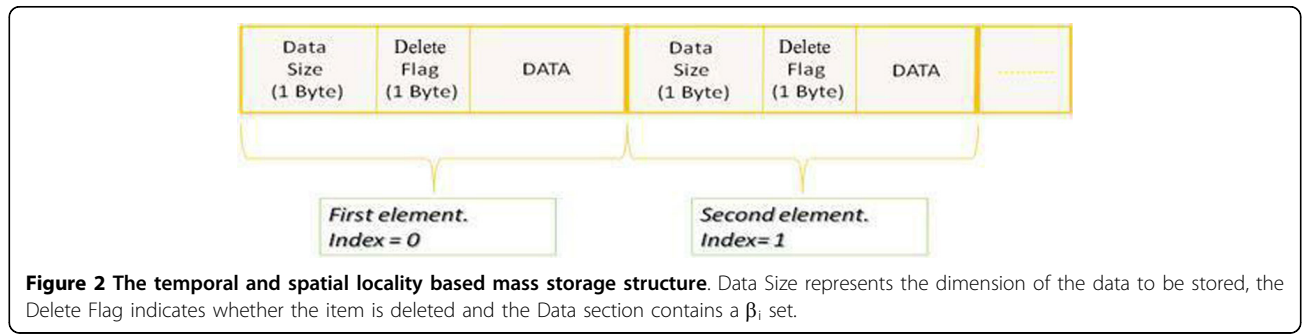

Figure 2 **The temporal and spatial locality based mass storage structure**. Data Size represents the dimension of the data to be stored, the Delete Flag indicates whether the item is deleted and the Data section contains a $\beta_i$ set.

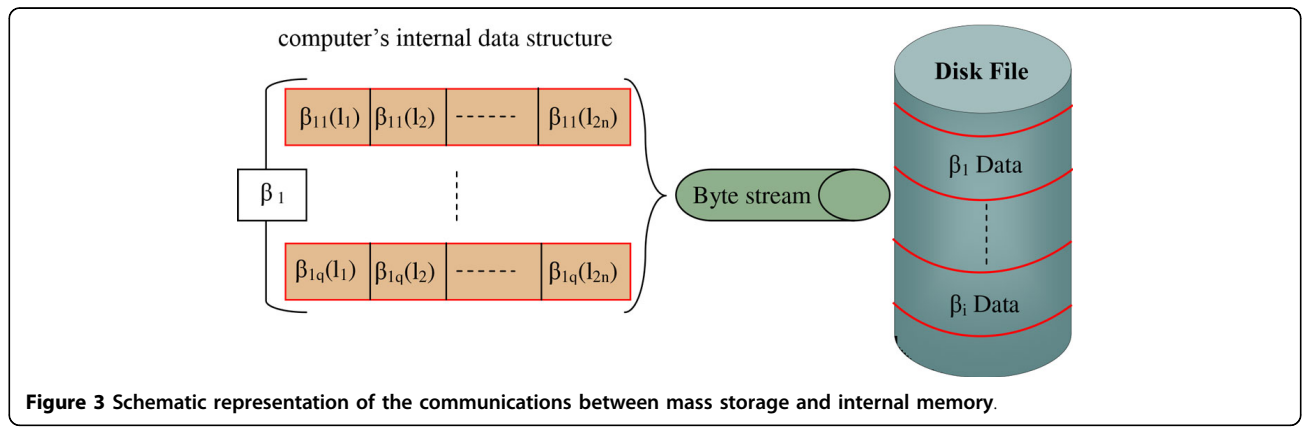

Figure 3 **Schematic representation of the communications between mass storage and internal memory**.

**Replacing $R_{ij}$ by $\beta_{ij}$**
In order to choose a literal with the highest relevance, the $R_i$ and $R$ vectors must be calculated in the inner cycle (2.5.) of the U-BRAIN algorithm. For each positive instance (fixed i-th index), the $R_i$ vector calculation requires the sum of the $R_{ij}$ relevances related to each of the negative instances (j ranging from 1 to q). This sum must be performed for each of the 2n literals. Thus, the $R_i$ vector calculation need two cycles, one on the negative instances and one on the 2n literals. However, the introduction of the $\beta_{ij}$ vectors has led to the following $R_i$ formula:

$$R_i(l_k) = \frac{1}{q} \Sigma_{j=1}^{q} \frac{\beta_{ij}^{\alpha}(l_k)}{\Sigma_{m=1}^{2n} \beta_{ij}^{\alpha}(l_m)} \qquad (19)$$

This allows to compute $R_i$ directly by $\beta_{ij}$. So the algorithm is further modified as follows:

1. Initialize f = Ø
2. While(∃ positive instances)
    2.1. Uncertainty Reduction
    2.2. Repetition Deletion
    2.3. Initialize term = Ø
    **2.4. Compute the $\beta_{ij}$ values**
    2.5. While(∃ elements in $\beta_{ij}$)
        2.5.6. Compute the $R_i$ relevances
        2.5.7. Compute the $R$ relevances
        2.5.8. Choose Literal
        2.5.9. Update term
        **2.5.10. Update $\beta_{ij}$**
    2.6. Add term to f
    2.7. Update positive instances
    2.8. Update negative instances
    2.9. Check consistency

By noting that the denominator in (19) does not change with the $l_k$ literals but only by j index, its value, for a fixed i index, can be calculated once and for all the 2n different literals $l_k$. In Figure 4 a comparison between the different methods to calculate $R_i$ for n = 2 is presented. The new manner of calculate the Ri, considers the rows of the table (right side in Figure 4) instead of the columns (left side in Figure 4). Thus, for each fixed j index the cardinality (denominator of $R_i$ formula) is calculated first and, then, the relevance on the 2n literals is carried out.

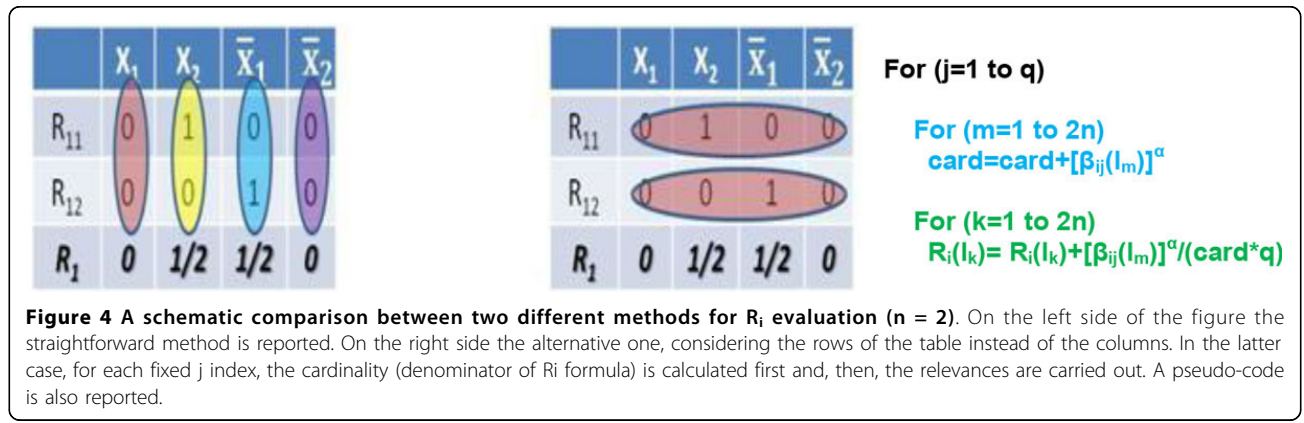

**Figure 4 A schematic comparison between two different methods for $R_i$ evaluation (n = 2)**. On the left side of the figure the straightforward method is reported. On the right side the alternative one, considering the rows of the table instead of the columns. In the latter case, for each fixed j index, the cardinality (denominator of Ri formula) is calculated first and, then, the relevances are carried out. A pseudo-code is also reported.

### Parallel programming model

Here we model a parallel implementation of the U-BRAIN algorithm.

There are three common strategies for creating parallel applications [44]. The first two, implicit parallelism, are based on the automatic parallelization of a sequential program, and on the use of parallel libraries that encapsulate some of the parallel code commonly used. The third one, explicit parallelism, involves the writing of the parallel application from the beginning. It was observed that the use of explicit parallelism, when properly applied, obtains a better efficiency than parallel language or compilers that use implicit parallelism [44]. This is the strategy we adopt here.

From a Flynn's taxonomy [45] point of view, we adopt a MIMD/Master-Slave strategy, and, specifically, a SPMD programming approach [46,47], together to the Message-Passing Programming Paradigm. In SPMD, multiple autonomous processors simultaneously execute the same program at independent points. That is, a single program is written so that different processes carry out different actions, and this is achieved by simply having the processes branch on the basis of their process rank. The Message-passing paradigm provides routines to initiate and configure the messaging environment, sending and receiving packets of data between processors of a parallel system.

The portability, the network transparency and the heterogeneity are other goals of interest. Currently, one of the most high-level message-passing systems is MPI (Message Passing Interface) defined by the MPI Forum [48]. MPI is a specification, not an implementation; there are multiple implementations of MPI including versions for COW (Cluster Of Workstation) [49], distributed-memory multiprocessors (MPP) and shared-memory machines (SMP). Here, we adopt a version for the Java programming language, MPJ [50]. Compared with C or Fortran, the advantages of the Java programming language include higher-level programming concepts, improved compile time and runtime checking, and, as a result, faster problem detection and debugging. In the context of "Java for HPC", the performance evaluation of the Java version reveals that it could achieve comparable performance to the original C code and the Java code performs better in the computation stages [51].

Although MPI offers great vantages, a significant amount of tasks of the parallelization are delegated to the programmer. So, a design methodology that allows the programmer to focus on machine-independent issues is desirable. According with Foster [52] we adopt a methodology organized into four stages: *partitioning, communication, agglomeration and mapping.*

### Partitioning

Each iteration of the U-BRAIN algorithm relies on the R relevance computation, that is:

$$R(l_k) = \frac{1}{pq} \Sigma_{i=1}^{p} \Sigma_{j=1}^{q} \frac{\beta_{ij}^{\alpha}(l_k)}{\Sigma_{m=1}^{2n} \beta_{ij}^{\alpha}(l_m)} \qquad (20)$$

$$= \frac{1}{p} \Sigma_{i=1}^{p} R_i(l_k) \qquad (21)$$

This computation can be easily decomposed into *nProc* tasks $\sigma_h$, where each task computes a partial summation as it follows:

$$R(l_k) = \Sigma_{h=1}^{nProc} \sigma_h \qquad (22)$$

$$= \frac{1}{p} \Sigma_{i=1}^{p1} R_i(l_k) + \frac{1}{p} \Sigma_{i=p1+1}^{p2} R_i(l_k) + ... + \frac{1}{p} \Sigma_{i=pn}^{p} R_i(l_k) \qquad (23)$$

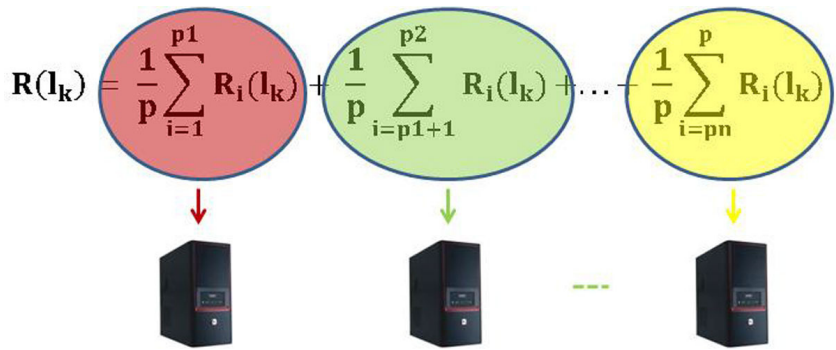

**Figure 5** Parallel implementation of U-BRAIN: each processor executes concurrently a partial summation.

Each sum treats some of the positive instances and all the given negative instances; $R_i$, indeed, is calculated by using all the q negative instances. This representation is a typical *domain decomposition* [52] that moves towards a parallel implementation where each process executes concurrently a partial summation (Figure 5).

The (22) leads to the implicit decomposition of the other tasks of the algorithm; in fact the uncertainty reduction, the repetition deletion and the updating of positives and negatives instances are performed concurrently by each process on their own partial data.

### Communication
The decomposition (22) of the U-BRAIN algorithm determines the need of a communication protocol among the processors involved; efficient synchronization design is essential. In our implementation each partial summation in (22) of R is performed by a single process (slave) which must communicate its own results to a unique master-process. The master process (also calculating a partial summation) waits the results from all the others processes and then it choose the term literal. Then, the master process sends the literal to all the processes involved. When the slave processes receive the literal, they continue their execution. When a negative instance is updated by a process, (2.7.), the new instance must be sent to all the processes. In this way, each process updates its own negative instance dataset in order to save the data consistency. All the processes receive a stop signal whenever an inconsistency issue arises. After each uncertainty reduction, repetition deletion or instance updating step, each process communicate its own results to all the others processes. So, for these tasks an all-gather technique [48] is used in order to implement a total exchange of data among the processes. A schematic representation of the communication protocol is shown in Figure 6.

This decomposition can be considered as the problem of performing a *parallel reduction operation*, that is, an operation that reduces nProc values, distributed over nProc tasks, at a single destination task using a commutative associative operator, in this case a summation. Because the master (RANK 0 process, see Figure 6) can receive and sum only one $R_i$ vector at a time, this approach takes O(nProc) time. A better performance could be obtained by using a divide and conquer strategy, commonly known as recursive doubling, that involves the splitting of the computation into pairs of sub-computation that can be performed concurrently. This approach would require O(log nProc) time. However the divide and conquer strategy introduces new communication and synchronization costs among the processes, especially when the process interconnection is made by switches. Figure 7 depicts an example of connection bottleneck; indeed, if a recursive doubling technique is used by coupling the processes 0-1, 2-3, 4-5, 6-7, each pair must wait the end of the communication between the previous pairs before starting.

### Agglomeration and mapping
In order to move the previous abstract phases toward the concrete implementation on a specific parallel system, we adapt the number of the partial summations in (23) to exactly one per processor. So, assuming to use nProc processors, each processor will treat, on average, $\left\lfloor \frac{p}{nProc} \right\rfloor$ positives instances. If p is not multiple of nProc, the division between p and nProc leads to a remainder (*p mod nProc*) different from zero. The remaining instances are distributed on *p mod nProc* processors that are charged of an additional task. In this case, our design is already largely complete, since in defining the nProc tasks that will execute on nProc processors, we have also addressed the mapping problem. This method of load balancing [53] is static because the tasks are assigned to the processors before the process starts and no information is collected about the state in real time of each single processor. Each processor, by acting on different positive instances, has different execution times. Thus it may finish the job and remains idle. In this case no further load balancing is made, so avoiding NP- Complete problems [54,55]. Nevertheless, the mapping between tasks and processors, used here, follows a semi-dynamic load balancing algorithm able to adapt the load to the number of available processors. Each process, on the basis of both its own identifier number (RANK) and the given number of positive instances (p), loads a fraction of the positive data. Each processor compute the $\beta_{ij}$ values and stores them in its own random access file according to the mass storage structure of Figure 2; in this way a reduction of data storage for each processor is also obtained and no file access synchronization is required.

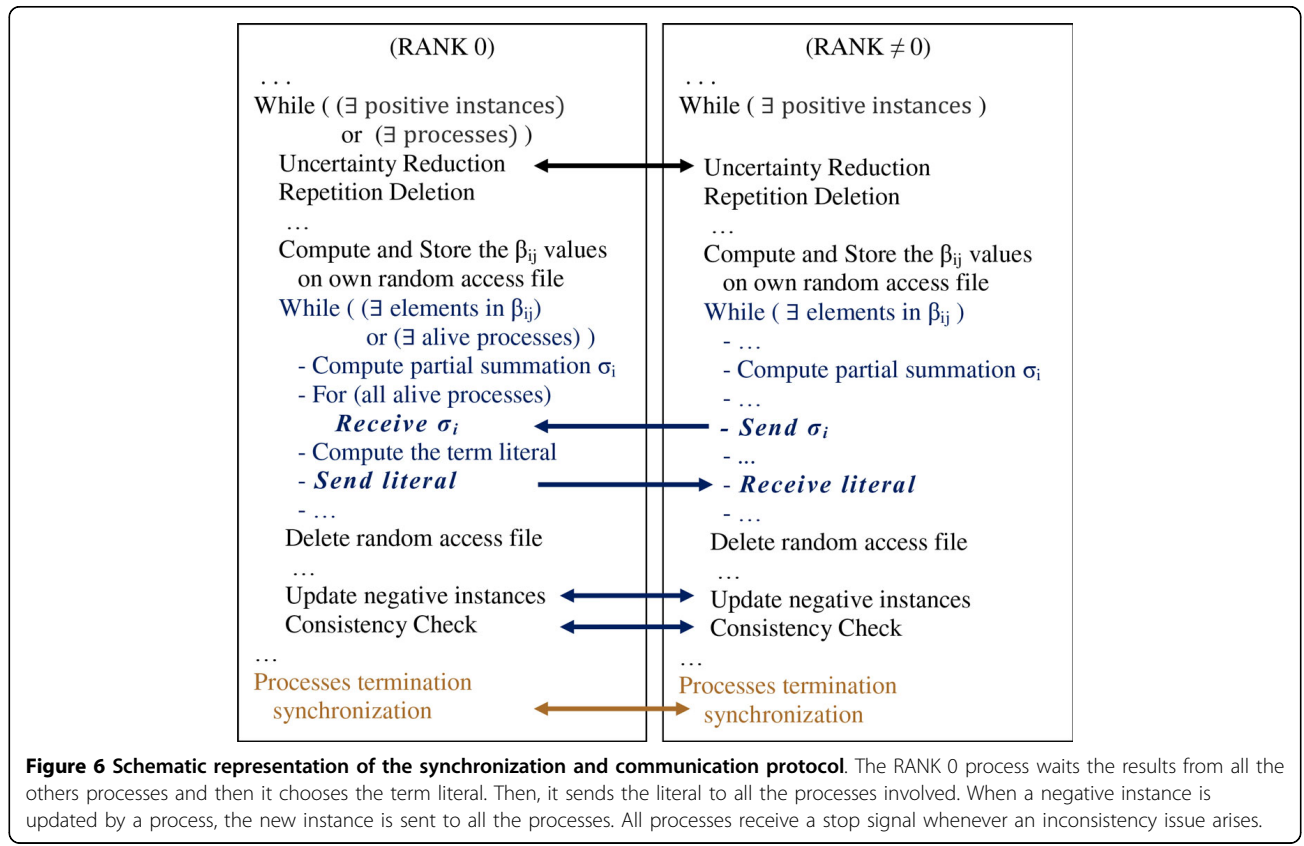

Figure 6 Schematic representation of the synchronization and communication protocol. The RANK 0 process waits the results from all the others processes and then it chooses the term literal. Then, it sends the literal to all the processes involved. When a negative instance is updated by a process, the new instance is sent to all the processes. All processes receive a stop signal whenever an inconsistency issue arises.

## Results and discussion

In the context of a collaboration between public and private institutions (Figure 8), the implementation has been tested on the INTEL XEON E7xxx and E5xxx family of the CRESCO structure of Italian National Agency for New Technologies, Energy and Sustainable Economic Development (ENEA), developed in the framework of the European Grid Infrastructure (EGI), a series of efforts to provide access to high- throughput

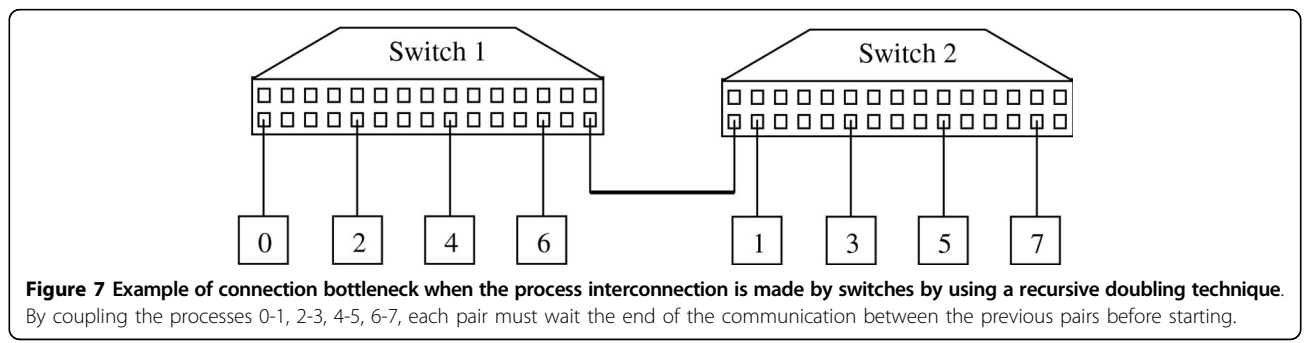

Figure 7 Example of connection bottleneck when the process interconnection is made by switches by using a recursive doubling technique. By coupling the processes 0-1, 2-3, 4-5, 6-7, each pair must wait the end of the communication between the previous pairs before starting.

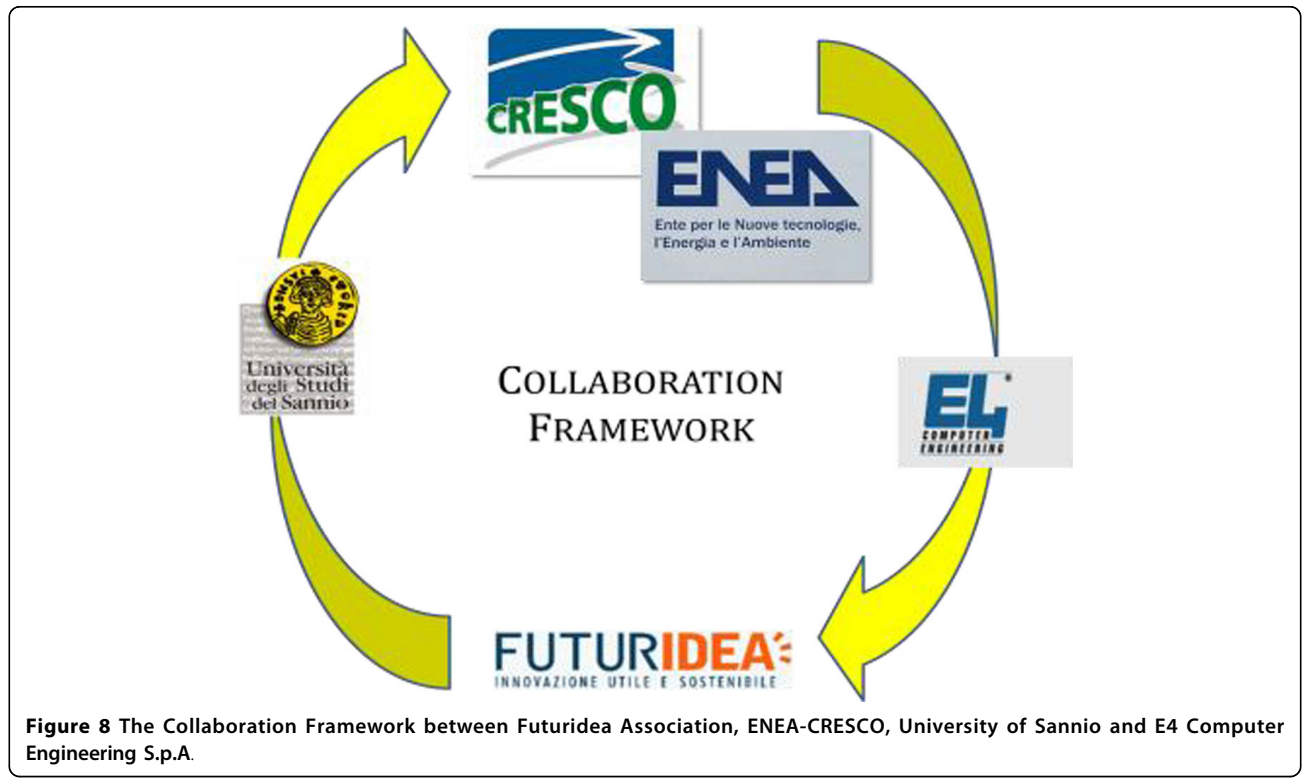

**Figure 8** The Collaboration Framework between Futuridea Association, ENEA-CRESCO, University of Sannio and E4 Computer Engineering S.p.A.

computing resources across Europe using grid computing techniques.

The computing resources and the related technical support used for this work have been provided by CRESCO/ENEAGRID High Performance Computing infrastructure and its staff.

### Cluster architecture

A schematic representation of the computer cluster used is shown in Figure 9. Such a distributed-memory system consists of a collection of core-memory pairs connected by a network, and the memory associated to a core is directly accessible only to that core. Each core is based on INTEL XEON E7xxx and E5xxx CPU family with a 2.40 GHz clock frequency. According to the specific cluster used (CRESCO1, CRESCO2, CRESCO3) the core memory size ranges from 16 GB to 64 GB, while the system storage is virtually unique, obtained by a GPFS distributed network file system. The communication network is based on multi-level Cisco switches on InfiniBand architecture.

### Data sets

The test data used in this study are IPDATA (Irvine Primate splice-junction data set) [56], a subset of HS3D (Homo Sapiens Splice Sites Dataset) [57,58] and a subset of COSMIC (the Catalogue of Somatic Mutations in Cancer) [59].

IPDATA is a data set of human splice sites, and it consists of 767 donor splice sites, 765 acceptor splice sites, and 1654 false splice sites. According to previous usage [30] we consider 464 positive instances and 1536 negative instances each one coded by 240 bits.

HS3D is a data set of Homo Sapiens Exon, Intron and Splice sites extracted from GenBank Rel.123. It includes 2796 + 2880 donor and acceptor sites, as windows of 140 nucleotides (560 bits) around a splice site, and 271,937 +332,296 windows of false splice sites, selected by searching canonical GT-AG pairs in not splicing positions. In this study we adopt a subset of 2974 donor sites and 161 false ones. COSMIC curates comprehensive information on somatic mutations in human cancer. Release v48 (July 2010) describes over 136,000 coding mutations in almost 542,000 tumour samples. Here we focus on the tumor suppressor *p16* [60]. 60 positive instances are generated from *missense* and *nonsense* mutations, while 62 negative instances are generated from *synonymous* mutations [60]. Each instance is of 1884 bits.

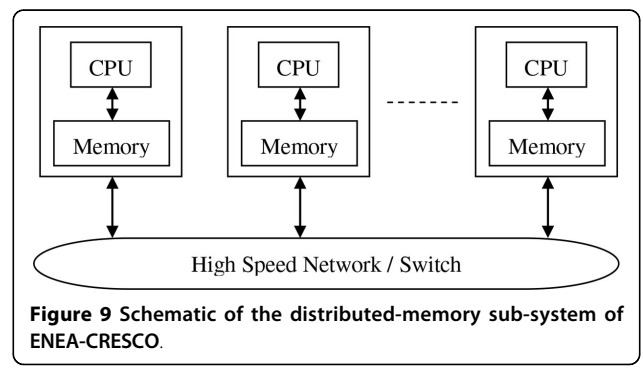

**Figure 9** Schematic of the distributed-memory sub-system of ENEA-CRESCO.

The dimension of each dataset used is estimated as p×q×n. In this way the size of HS3D sub-dataset is greater than IPDATA which is greater than *p16* COSMIC.

### Experimental results

Speed-up is the most basic methods to evaluate the performance of a parallel program [61]. Speed-up refers to how much the parallel program is faster than the corresponding sequential one. It is defined as follows:

$$S(nProc) = \frac{T_S}{T_{nProc}} \quad (24)$$

where *nProc* is the number of processors, $T_S$ is the execution time of the sequential program and $T_{nProc}$ is the execution time of the parallel one with *nProc* processors. Ideal speed-up is obtained when $S(nProc) = nProc$, while for one processor $S(1) = 1$.

An estimate of this parameter has been taken into account during all the design and the testing of the parallelization process. The use of several data sets has been useful in order to show the effect of the *granularity* on the performance varying the problem size. The granularity is a qualitative measure of the ratio of computation to communication [61]. Two graphs, reporting the execution time and the speed-up of the parallel version, respectively, are shown for each data set.

The execution times of the parallel implementation varying the number of processors on IPDATA are reported in Figure 10, while the speed-up of the parallel implementation varying the number of processors on IPDATA is reported in Figure 11.

As evidenced in Figure 10 and 11, the execution times and the speed-up on IPDATA reach the best values within about 90 processors. Then the parallelization advantage is balanced by the greater cost of non-local communications between the processors.

As shown in Figure 12 and 13, a similar behavior is evidenced by running the program on HS3D, but at a greater number of processors; this confirms the direct relationship between data size and parallelization gain.

The results on COSMIC, reported in Figure 14 and 15, evidence that, in this case, as the number of processors grows, the workload of the processors remains higher than the cost of the non-local communications. In the Figures the maximum number of processing elements is 60, since it is upper bounded by the number of positive examples p.

Overall, the results obtained on the data sets used show that the parallel version is up to 30 times faster than the serial. Moreover, increasing the problem size, at constant number of processors, the speed-up averagely increases.

### Conclusions

High-throughput technologies are producing an increasing amount of experimental and clinical data. In such a scenario, large-scale databases and bioinformatics methods are key tools for organizing and exploring biological

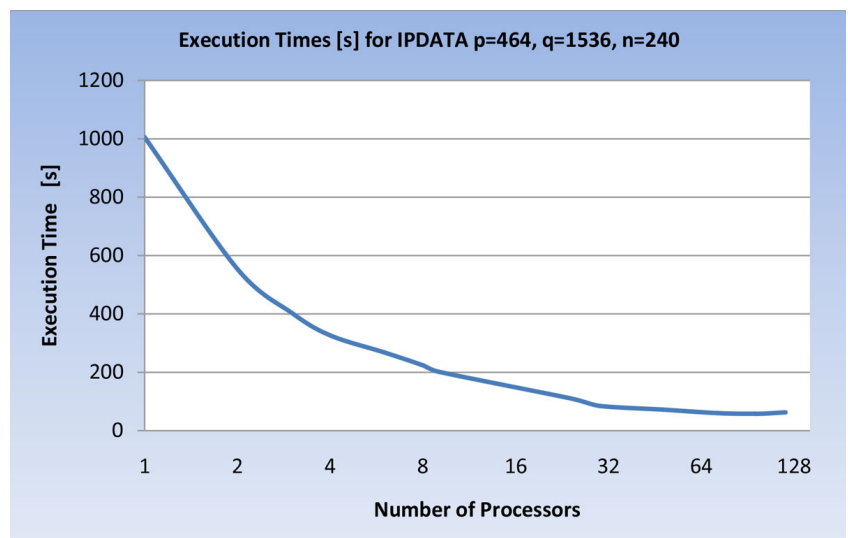

**Figure 10 Execution times, of the U-BRAIN parallel implementation on IPDATA varying the processor number (log scale)**.

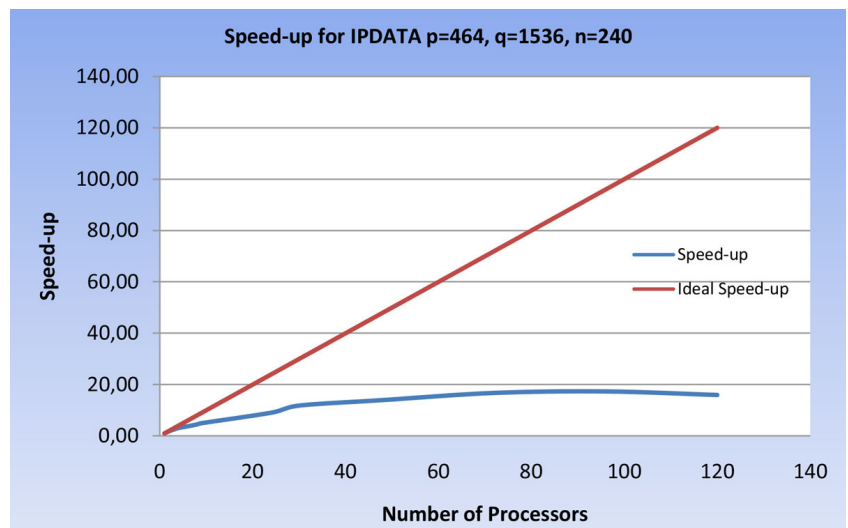

**Figure 11 Speed-up of the U-BRAIN parallel implementation on IPDATA varying the processor number**.

and biomedical data with the aim to discover new knowledge in biology and medicine.

High-performance computing may play an important role in many phases of life sciences research, from raw data management and processing, to data analysis and integration, till data exploration and visualization. In particular, at the raw data layer, Grid infrastructures may offer the huge data storage needed to store experimental and biomedical data, while parallel computing can be used for basic pre-processing and for more advanced analysis. In such a scenario, parallel architectures coupled with specific programming models may overcome the limits posed by conventional computers to the mining and exploration of large amounts of data.

Here we investigated the problems arising from the HPC implementation of a general purpose learning algorithm able to treat uncertainty on data (U-BRAIN). The U-BRAIN algorithm can be used in many fields of the biology in order to extract the laws that govern the biological process, in the form of mathematical formulas. The U-BRAIN parallel implementation aims to override the computational limits that make the algorithm unaffordable for huge data sets. We found mathematical and programming solutions able to effectively implement the algorithm U-BRAIN on parallel grid computers. The implementation is able to minimize both the memory space and the execution time, while maintaining the results of the sequential version. The solutions adopted in this paper, e.g. dynamic programming, data representation minimization, efficient use of memory, mass storage unit structure with a high degree of temporal and spatial locality, SPMD parallel implementation and Message-Passing Programming Paradigm, are tailored for the U-BRAIN algorithm, but they can be used for many others HPC-oriented parallel implementations.

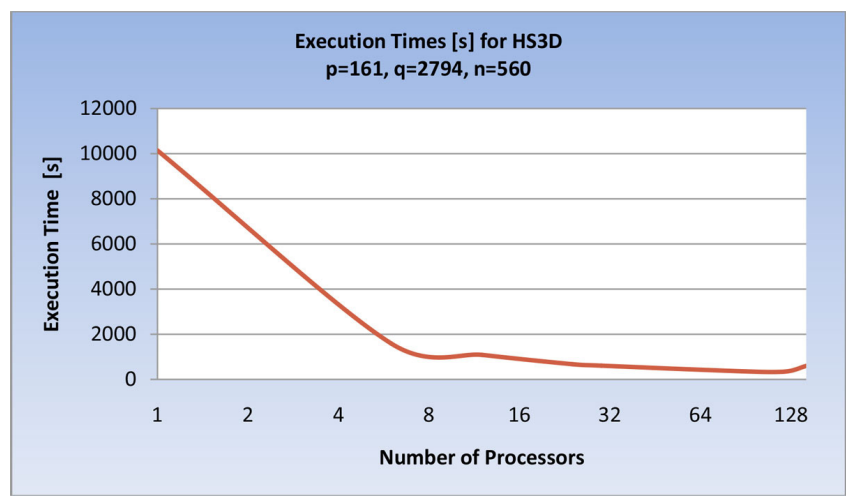

**Figure 12 Execution times, of the U-BRAIN parallel implementation on HS3D varying the processor number (log scale)**.

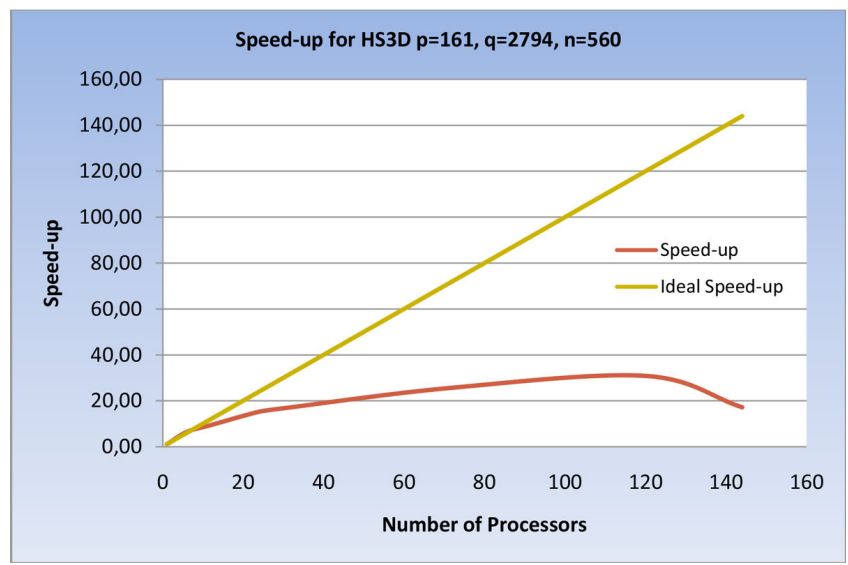

Figure 13 Speed-up of the U-BRAIN parallel implementation on HS3D varying the processor number.

As evidenced in the experiments, the execution times and the speed-up reach the best values within a data dependent number of processors. Then the parallelization advantage is balanced by the greater cost of non-local communications between the processors. This evidences the direct relationship between data size and parallelization gain.

The obtained results, though not excellent in terms of performance, encourage the algorithm application on larger data sets. By applying the U-BRAIN algorithm on the full HS3D data set (p = 2796, q = 271937, n = 560), and using a single processor, $0,17 \times 10^8$ seconds (197 days) are needed to reach the result, while, by assuming a linear relation among the dimension and the execution time, the time reduces to $0,57 \times 10^6$ seconds (about 6 days) in a parallel configuration with speed-up = 30.

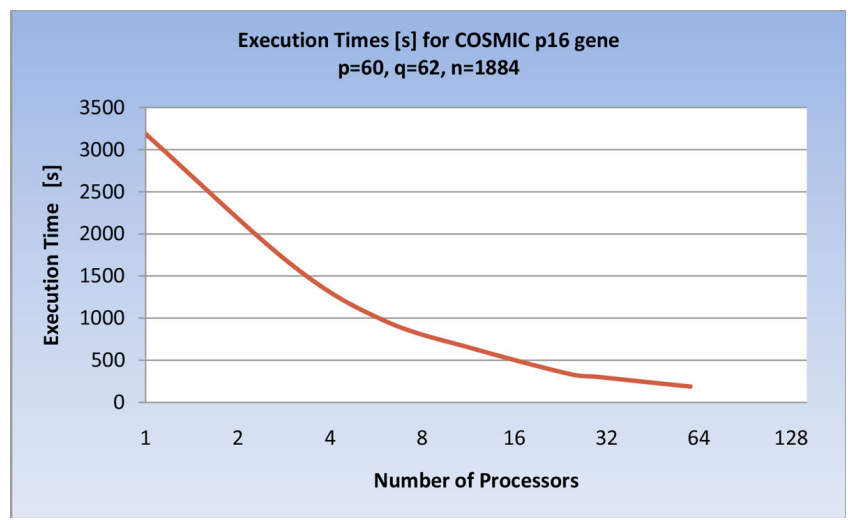

Figure 14 Execution times of the U-BRAIN parallel implementation on COSMIC *p16* gene varying the processor number (log scale).

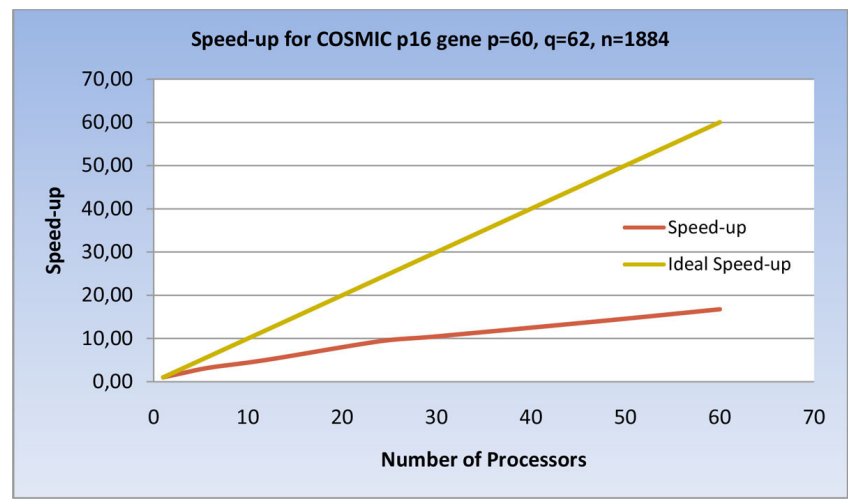

**Figure 15 Speed-up of the U-BRAIN parallel implementation on COSMIC *p16* gene varying the processor number**.

Open problems rest in adopting a dynamic load balancing algorithm, capable of migrating the load among the processors. Load balancing problem is a most critical point in parallel computing design [53]. A more thorough assessment of how the synchronization and communication costs affect the total performance varying the problem size is another issue that we refer to future works. In order to increase the performance it would be useful implement U-BRAIN algorithm by using a hybrid MPI/OpenMP programming on clusters of multi-core with shared-memory nodes [62].

#### Competing interests
The authors declare that they have no competing interests.

#### Authors' contributions
SR conceived of the project idea. Both authors contributed to the implementation design. GDA implemented the algorithm on the CRESCO structure, and performed the simulations. Both authors contributed to the test cases analysis and interpretation, and wrote the final manuscript.

#### Acknowledgements
This work was supported by the collaboration agreement between Futuridea Association, ENEA, University of Sannio and E4 Computer Engineering S.p.A. The authors wish to thank the CRESCO/ENEAGRID High Performance Computing staff and in particular Ing. Silvio Migliori, Ing. Antonio Perozziello, and Ing. Guido Guarnieri. CRESCO/ENEAGRID High Performance Computing infrastructure is funded by ENEA, the Italian National Agency for New Technologies, Energy and Sustainable Economic Development and by national and European research programs.

#### Declarations
The publication costs for this article were funded by DST University of Sannio.
This article has been published as part of *BMC Bioinformatics* Volume 15 Supplement 5, 2014: Italian Society of Bioinformatics (BITS): Annual Meeting 2013. The full contents of the supplement are available online at http://www.biomedcentral.com/bmcbioinformatics/supplements/15/S5

#### Authors' details
[1]Department of Science and Technology (DST), University of Sannio, Benevento, Italy. [2]Futuridea Innovazione Utile e Sostenibile, Benevento, Italy.

Published: 6 May 2014